\documentclass[10pt,a4paper,twoside]{article}
\usepackage{epsfig}
\usepackage{baltlat6}
\usepackage{array}
\usepackage{here}
\begin{document}
\ \
\vspace{0.5mm}
\setcounter{page}{137}

\titlehead{Baltic Astronomy, vol.\,24, 137--143, 2015}

\titleb{DETECTION OF UNRESOLVED BINARIES WITH MULTICOLOR PHOTOMETRY}

\begin{authorl}
\authorb{D.~Chulkov}{1},
\authorb{M.~Prokhorov}{2},
\authorb{O.~Malkov}{1},
\authorb{S.~Sichevskij}{1},
\authorb{N.~Krussanova}{2},\\
\authorb{A.~Mironov}{2},
\authorb{A.~Zakharov}{2} and
\authorb{A.~Kniazev}{2,3,4}
\end{authorl}

\begin{addressl}
\addressb{1}
{Institute of Astronomy of the Russian Academy of Sciences, Pyatnitskaya St.\\
48, Moscow 119017, Russia; chulkov@inasan.ru}%, malkov@inasan.ru, s.sichevskij@gmail.com}
\addressb{2}
{Sternberg Astronomical Institute, M. V. Lomonosov Moscow State University,\\
Universitetskij Prosp. 13, Moscow 119991, Russia;
mike.prokhorov@gmail.com}
 \addressb{3}{South African Astronomical Observatory, PO Box 9,
Observatory, Cape Town 7935, South Africa; akniazev@saao.ac.za}
\addressb{4}{South African Large Telescope Foundation, PO Box 9,
Observatory, Cape Town 7935, South Africa}
\end{addressl}

\submitb{Received: 2015 March 25; accepted: 2015 April 20}

\begin{summary} The principal goal of this paper is to specify
conditions of detection of unresolved binaries by multicolor
photometry.  We have developed a method for estimating the critical
distance at which an unresolved binary of given mass and age can be
detected.  The method is applied to the photometric system of the
planned {\it Lyra-B} spaceborne experiment.  We have shown that some
types of unresolved binary stars can be discovered and distinguished
from single stars solely by means of photometric observations.
\end{summary}


\resthead {Detection of unresolved binaries with multicolor photometry}
{D. Chulkov, M.~Prokhorov, O.~Malkov et al.}

\sectionb{1}{INTRODUCTION}

Binary stars are very numerous, however only the nearest pairs can
be resolved into components.  Others remain unresolved, and the vast
majority of them are observed neither as spectroscopic nor as
eclipsing binaries.  On the other hand, the effect of unresolved
binaries is extremely important for the construction of the
luminosity function (Piskunov \& Malkov 1991; Kroupa et al. 1991)
and the initial mass function (Malkov 2002). Therefore, the
detection and analysis of photometrically unresolved binaries remain
a vital and promising avenue of investigation.

If both components of a binary star have relatively different
effective temperatures, but comparable luminosities, the color
indices of such a system will differ from those of a single star.
Generally speaking, even a binary with identical components can be
distinguished from single stars, because it is located on the HR
diagram above the position of the corresponding single star.
However, in many cases we do not have reliable estimates of stellar
distances, therefore it is usually impossible to distinguish between
a binary and a single star solely by means of one-band photometry.

The problem of detection and classification of unresolved binaries
by multicolor photometry was studied for the {\it Gaia} (Malkov et
al. 2011a) and ultraviolet (Malkov et al. 2011b) photometric
systems. Also, various methods were proposed for the identification
of unresolved binaries among very low mass stars and brown dwarfs by
spectroscopic observations (Bardalez Gagliuffi et al. 2014) or by
SDSS/2MASS cross-matching (Geissler et al. 2011).  In the current
paper we develop a new method for binary detection, estimate the
critical distance at which binary stars can be discovered, and apply
the method to the {\it Lyra-B} photometric system.

%%%%%%%%%%%%%%%%%%%%%%%%%%%%%%%%%%%%  FIGURE 1

\begin{figure}
\centerline{\includegraphics[width=7.8cm]{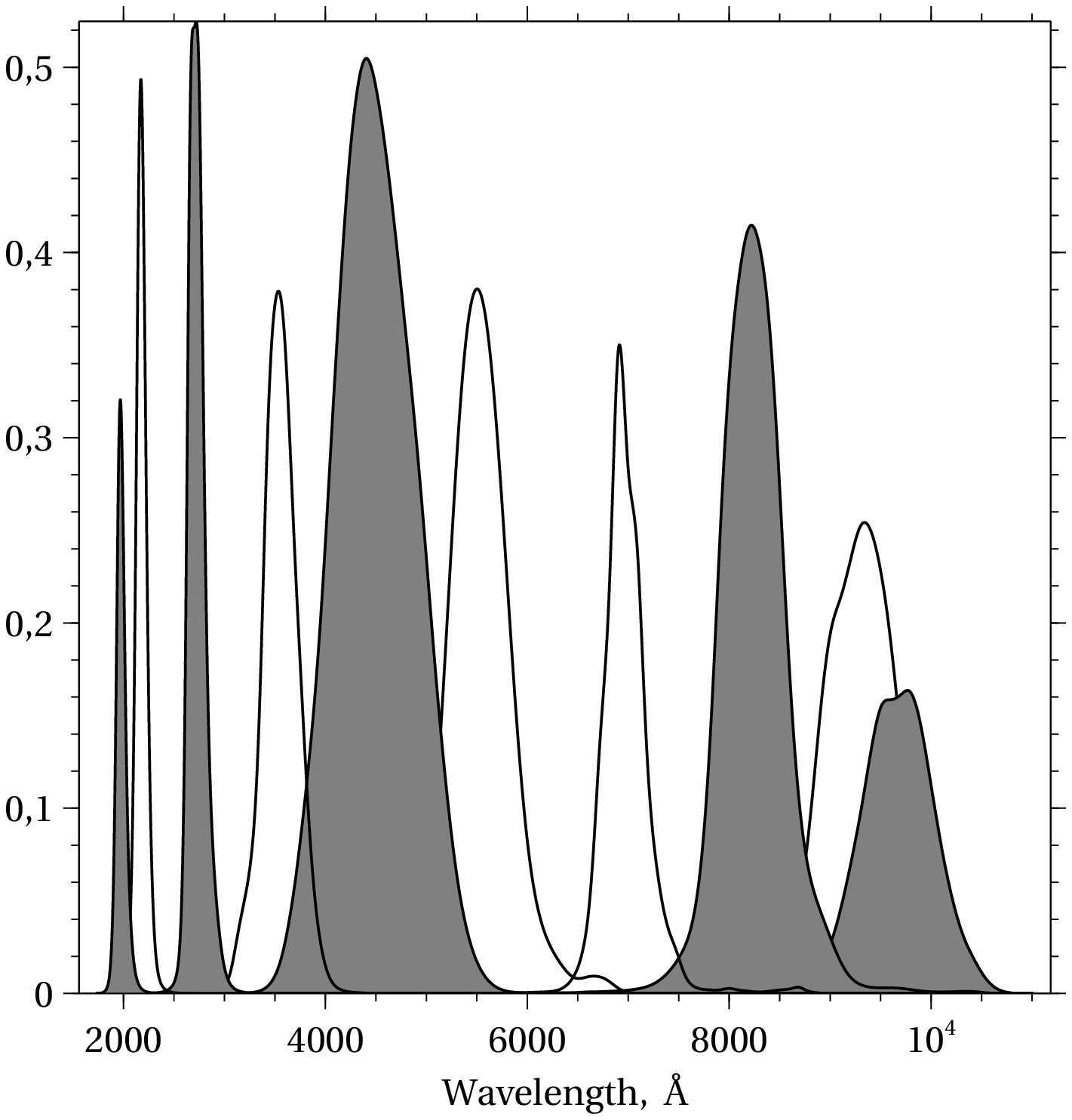}}
\captionb{1}{Response curves of the {\it Lyra-B} photometric system.
Transmission of the interference filters and CCD quantum efficiency
are taken into account.}
\end{figure}

{\it Lyra-B} (Zakharov et al. 2013a) is a space-based experiment
onboard the International Space Station, which is currently being
prepared at the Sternberg Astronomical Institute.  The main goal of
the experiment is to carry out a high-accuracy, multicolor all-sky
survey of stars down to 16--17 mag.  The response curves of the {\it
Lyra-B} ten-band photometric system are shown in Fig.~1, and the
central wavelengths of the bands are 195, 218, 270, 350, 440, 555,
700, 825, 930, 1000 nm.  During the {\it Lyra-B} mission, an all-sky
high-precision photometric survey of objects from 3 to 16 mag will
be undertaken.  The photometric errors of the survey catalog are
expected to be about 0.001--0.003 mag for stars brighter than 12 mag
and about 0.01 mag for fainter stars.  The estimated duration of the
survey is about five years. The {\it Lyra-B} photometric system is
described in detail by Zakharov et al.  (2013b), while the expected
characteristics of output data of the mission can be found in
Zakharov et al.  (2013c).

The proposed method for the detection of unresolved binaries is
described in Section 2. The application of the method to the {\it
Lyra-B} photometric system is illustrated in Section~3.  In
Section~4 we draw our conclusions.

\sectionb{2}{DESCRIPTION OF THE METHOD }

Theoretical color indices of binary stars can be obtained from
tracks and isochrones of single stars (this provides us with stellar
astrophysical parameters:  bolometric luminosity $L$, effective
temperature $T_{\rm eff}$ and surface gravity $g$, for given mass
$M$, age $\tau$ and metallicity) and with grids of model atmospheres
(which simulate observational photometry for a star with given
astrophysical parameters).  Then, the simulated fluxes, when
combined with the photometric system's response curves, provide us
with theoretical color-indices. The problem of the accuracy of
astrophysical stellar parameters determined from multicolor
photometry was studied by Sichevskij et al.  (2014).

To generate isochrones, we used PARSEC stellar evolution code
(Bressan et al. 2012) for the age $\tau = 10^8$ years and solar
metallicity. The astrophysical parameters $L$, $T_{\rm eff}$ and $g$
were obtained for stellar masses in the range 0.1--5.3~$M_\odot$,
with a step of $0.1 M_\odot$.  The upper mass limit is determined by
the age value ($\tau = 10^8$ yr).  We used a linear approximation to
compute the astrophysical parameters for a star of a particular
mass.

To get the fluxes for single stars we then used grids of ATLAS9
model atmospheres (Castelli \& Kurucz 2003) for $T_{\rm eff}$, $\log
g$ and interstellar extinction $A_V$, calculated with steps of 250K,
0.25 dex and 0.25 mag, respectively.  Throughout the paper, we adopt
the total to selective extinction ratio $R_V=3.1$.  We approximate
the flux of any given intermediate point ($T_{\rm eff}$, $\log g$,
$A_V$) within the local axial rectangular prism by trilinear
interpolation, using data at the eight corner points (reference
stars with known fluxes) on the cube surrounding the interpolation
point.  The normalized flux is considered to be $F(\lambda =
550\,{\rm nm}) = 1$.

%%%%%%%%%%%%%%%%%%%%%%%%%%%%%%  FIGURE 2

\begin{figure}
\centerline{\includegraphics[width=7.8cm]{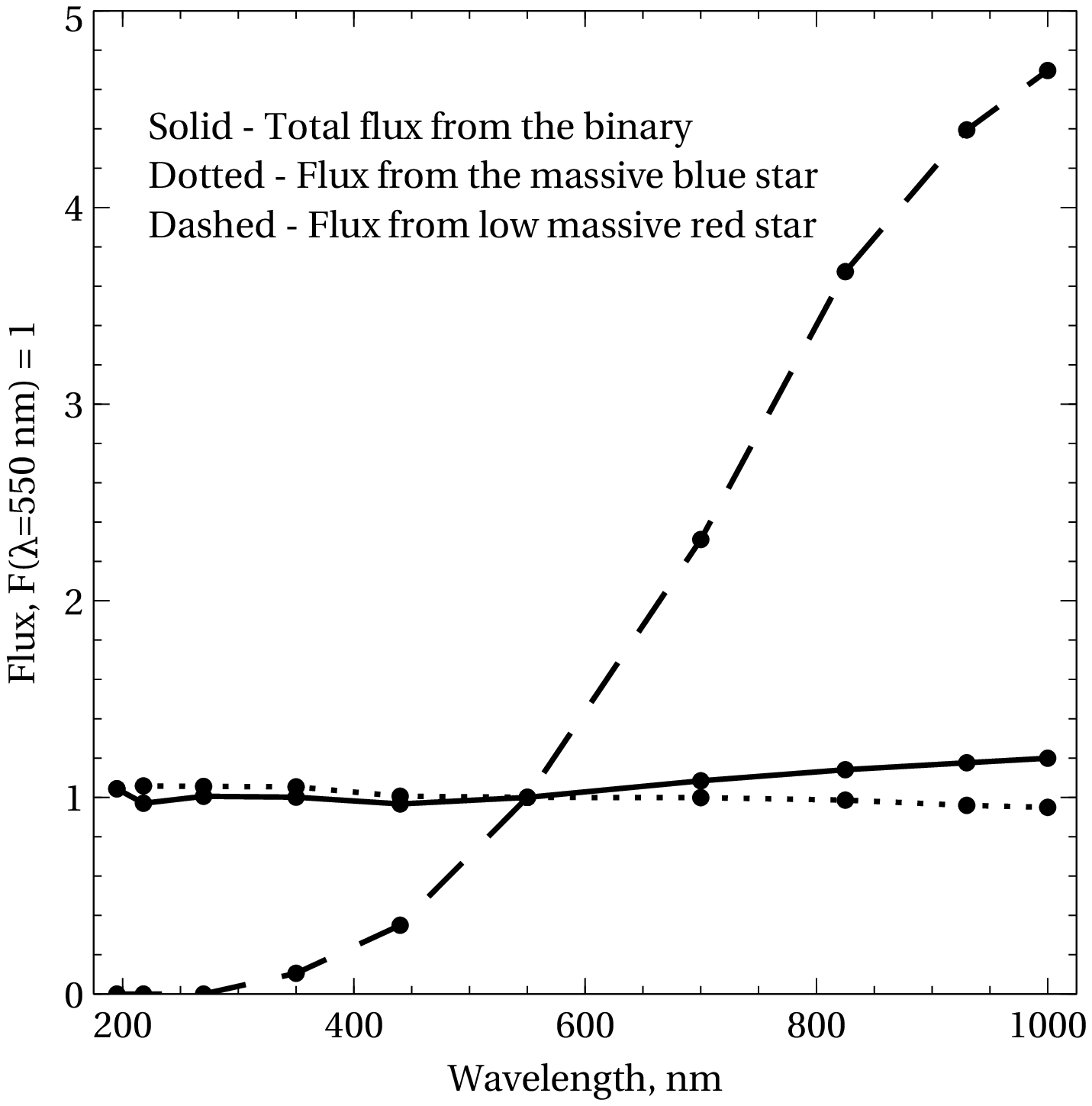}}
\captionb{2}{Example of spectral energy distribution (SED) for a
binary star. The dashed and dotted lines are SEDs of the components,
while the solid line represents the total flux.}
\end{figure}

The fluxes of two components are combined in accordance with their
luminosity ratio (see Fig.~2), and the color indices are calculated
for given response curves.
% Second component is selected from the list of masses.
We assume extinction $A_V$ to be the same for both components. The
resulting color indices of the binary star are compared with those
of single stars from the atlas, for various $A_V$ values.  The
best-fit single star is selected via the $\chi^2$ minimization
procedure:

\begin{equation}
\chi_{0}^{2} = \sum\limits_{k}^{1..10} \frac{(f_{\rm b}(k)-f_{\rm
s}(k))^2}{f_{\rm b}(k)},
\end{equation}
where $k$ is spectral band number, $f_{\rm b}(k)$ and $f_{\rm s}(k)$
are the binary and single comparison star fluxes in the given band
$k$, respectively.

The normalized fluxes are considered to be $f_{\rm
s}(\lambda=550)=f_{\rm b}(\lambda=550)=1$.  For a real star, $f_{\rm
b}(\lambda=550)\neq1$, and $\chi^2_{\rm obs} = \chi_{0}^{2}\cdot N
(\lambda=550)$.  Thus we find a critical distance $d_{\rm crit}$,
for which $\chi^2_{\rm obs}$ equals $\chi_{\rm crit}^2$.  Critical
$\chi_{\rm crit}^2$ is selected manually, and in our calculations we
assumed that $\chi_{\rm crit}^2=27$, which corresponds to a
confidence level of more than 99\%.

Note that in this paper we refer exclusively to non-interacting
binaries.

\sectionb{3}{RESULTS}

Hereafter, we consider non-reddened binary systems. Our estimates
show that for reddened systems, at least up to $A_V$ = 2 mag, the
results are qualitatively the same, though all the resulting values
of $d_{\rm crit}$ should be decreased.  However, we study the effect
of reddening of comparison single stars.  In most of the cases
considered, the best-fit single star is not reddened. Interstellar
reddening can be large enough for well evolved massive primaries
($A_V$ up to 2.7 mag), but for main-sequence stars the largest
reddening, $A_V$ = 0.44 mag, is achieved just for the
$0.7\,M_{\odot}+0.4\,M_{\odot}$ system.  In the latter case, the
single comparison star is hotter (4600K) than both the primary
(4500K) and the secondary (3900K).  The temperature of the
comparison star is usually intermediate between those of the two
components.

%%%%%%%%%%%%%%%%%%%%%%%%%%%%%%%  FIGURE 3

\begin{figure}
\centerline{\includegraphics[width=7.8cm]{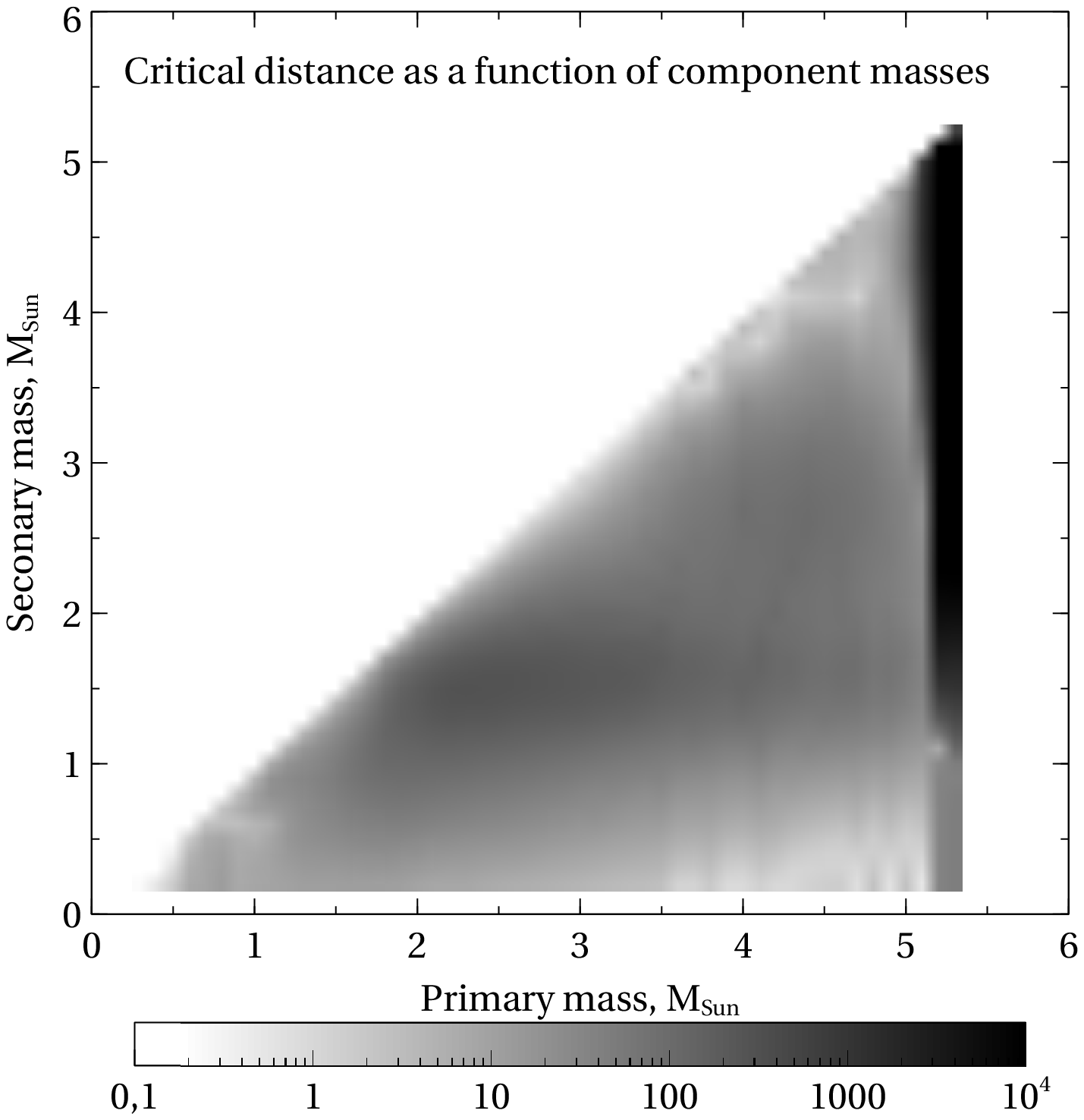}} \captionb{3}
{Critical distance of binary detection as a function of the
component masses. The darker shading corresponds to larger
distances.}
\end{figure}

%%%%%%%%%%%%%%%%%%%%%%%%%%%%%%%%  FIGURE 4

\begin{figure}
\centerline{\includegraphics[width=7.8cm]{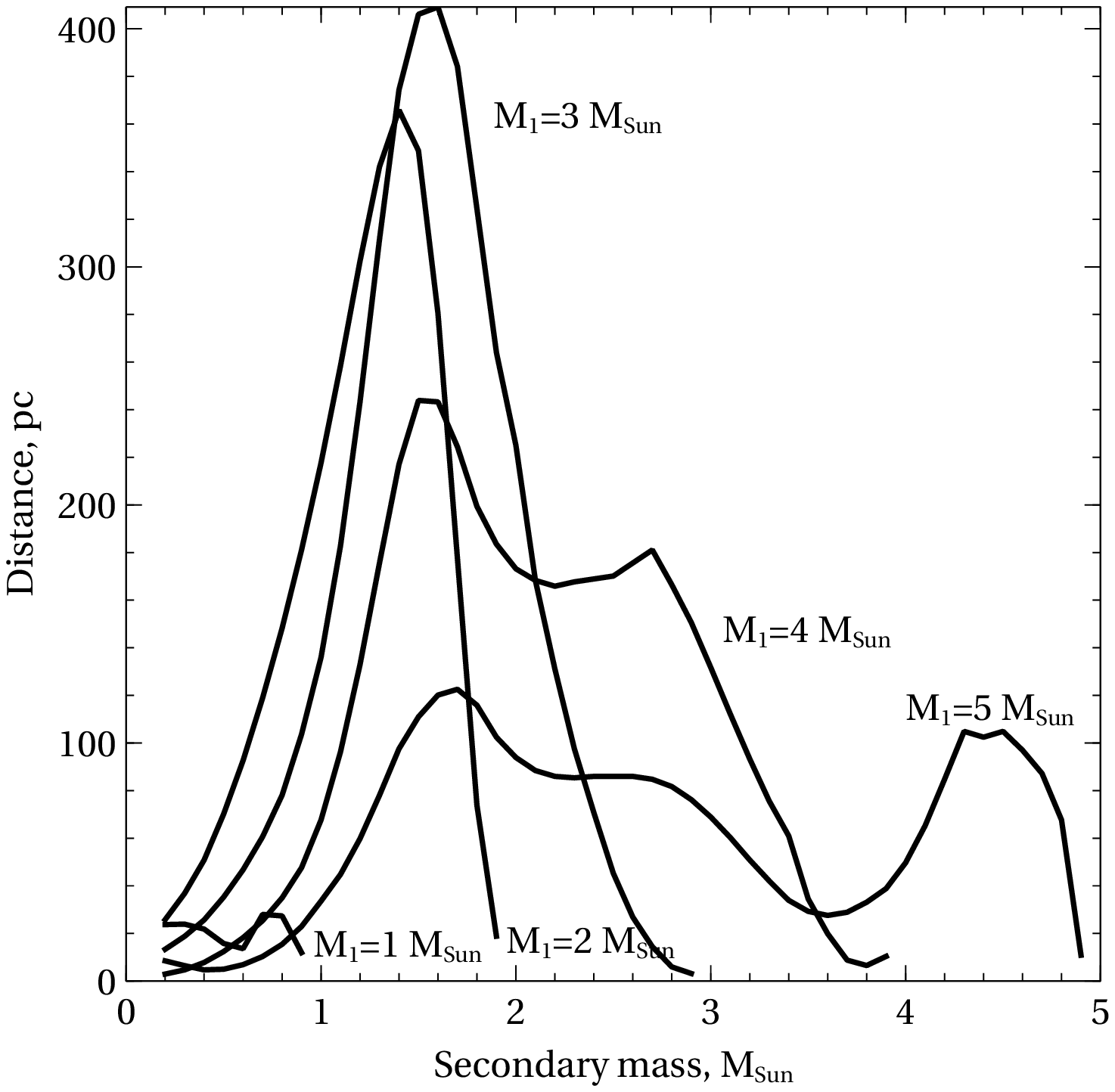}} \captionb{4}
{Critical distance of binary detection as a function of the
secondary mass for primary masses $M_1$ = 1, 2, 3, 4 and 5
$M_\odot$.}
\end{figure}

Fig.~3 shows the critical distances estimated for the isochrone
$\tau = 10^8$ years.  For further analysis we fix the primary mass
($M_1$) and estimate the critical distance as a function of the mass
of the secondary.  Fig.~4 shows the results for $M_1$ = 1, 2, 3, 4
and 5 $M_\odot$.

Because of the low luminosity of subsolar-mass binaries, their critical
distance does not exceed $d_{\rm crit} \approx 35$ pc.  Pairs with low
mass ratios $q=M_2/M_1$ tend to have better chances to be detected as
binaries.

The distribution curves for the primary masses 2 and 3 $M_\odot$
show a well pronounced single peak.  Stars with similar masses have
very similar spectra, making it hard to detect their binarity from
multicolor photometry.  On the other hand, pairs with a low-mass
companion contribute negligibly to the part of the distribution
corresponding to distinctly red secondaries because of the
dramatically low luminosity of the former.  These two phenomena
produce a steep peak.  According to our calculations, the maximum
critical distance is achieved for a $2.3\,M_{\odot} +
1.5\,M_{\odot}$ binary star, which can be detected out to a distance
of 500 pc.

For more massive primaries, the peak is less pronounced, and turns
into a plateau.  In particular, for a 4\,$M_\odot$ primary, the
distribution shows two extrema, at the secondary masses 1.5$M_\odot$
and 2.7\,$M_\odot$.  This can be explained by a greater variety of
single stars used for the comparison.  Consequently, it is much
easier to find a single star emulating a particular binary, and this
causes some drop in the critical distance (as it can be seen in
Fig.~4).  The maximum critical distance drops down to 325 and 175 pc
for the 3.5\,$M_\odot$ and 4.5$M_\odot$ primary, respectively.

%%%%%%%%%%%%%%%%%%%%%%%%%%%%  FIGURE 5

\begin{figure}
\centerline{\includegraphics[width=7.8cm]{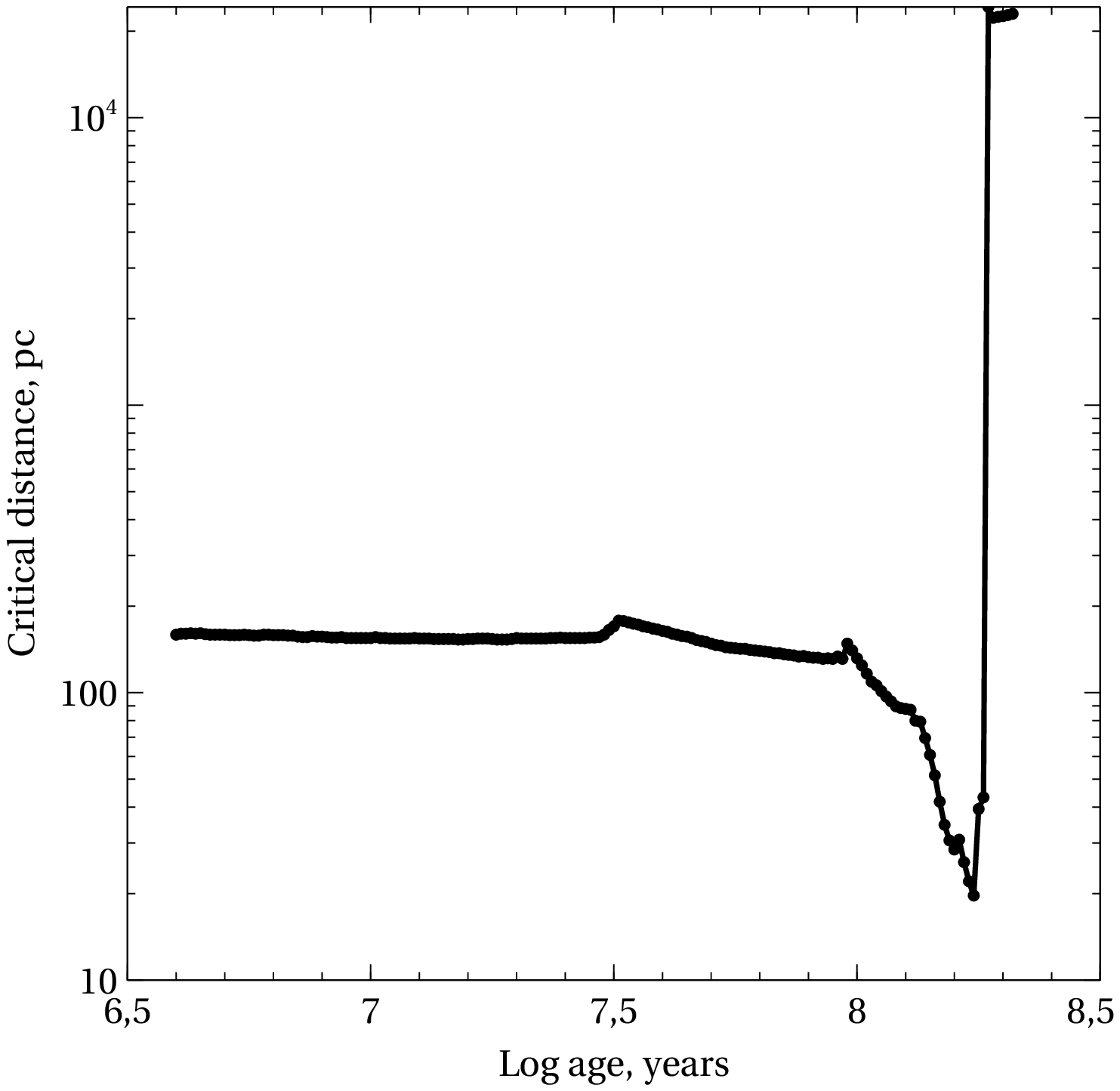}} \captionb{5}
{Critical distance of binary detection as a function of age for the
system with a 4\,$M_\odot$ primary and a 3\,$M_\odot$ secondary.}
\end{figure}

The calculations discussed above are obtained for a single
isochrone, and we therefore need to investigate the evolutionary
effects.  Because the calculations are time consuming we focus only
on a particular binary and calculate its detectability throughout
the lifetime.  The binary considered consists of a 4\,$M_\odot$
primary and a 3\,$M_\odot$ secondary.  Given an upper limit for the
lifetime of the primary, we performed our calculations for the ages
ranging from $\log \tau = 6.6$ (around 4 million years) to $\log
\tau = 8.32$ (around 200 million years), with a step of $\log \tau =
0.01$. The results are shown in Fig.~5.

Initially, the primary star has a higher temperature and luminosity
than its companion.  In the process of stellar evolution the
luminosities of both components gradually increase and their
effective temperatures decrease.  While on the main sequence, this
process is slow and has no significant effect on the observed color
indices of the system.  This means that our above predictions for a
certain isochrone apply to the entire main-sequence stage.  This
period lasts until about $\log \tau \approx 8.0$ (100 million years)
for the binary considered.  Of course, it would last much longer for
less massive and more numerous solar-mass primaries.

After the primary component leaves the main sequence, its evolution
speeds up and evolutionary processes occur on a shorter timescale.
The temperature decreases more rapidly and, at some point (around
$\log \tau \approx 8.23$ in our case), it reaches that of the less
massive companion, which is still on the main sequence.  This makes
the binary detection much more difficult, and the critical distance
reaches its minimum value.  Finally, the primary reaches the
Asymptotic Giant Branch, and we are left with a system of markedly
different (bright red primary and blue secondary) components.  The
difference of the effective temperatures becomes by far the largest
during the lifetime of the system, enabling the best conditions for
multicolor observations.

\sectionb{4}{CONCLUSIONS}

We developed a method for detecting unresolved binaries based on
multicolor photometry.  We apply the method to the ten-band
photometric system of the planned {\it Lyra-B} mission.  We estimate
the critical distance for the detection of a binary as a function of
its age and component masses.  Our method can also be extrapolated
to higher masses and can also be applied for other photometric
systems.

\thanks{ The work was partly supported by the Program of fundamental
research of the Presidium of the Russian Academy of Sciences (P-41),
the Russian Foundation for Basic Research (project No.~15-02-04053),
and the Program of support of leading scientific schools of the
Russian Federation (3620.2014.2).  A. K. acknowledges support from
the National Research Foundation of South Africa.}

\References

\refb Bardalez Gagliuffi D. C., Burgasser A. J., Gelino Ch.  R. 2014,
ApJ, 794, 143

\refb Geissler K., Metchev S., Kirkpatrick J. D. 2011, ApJ, 732, 56

\refb Bressan A., Marigo P., Girardi L. et al. 2012, MNRAS, 427, 127

\refb Castelli F., Kurucz R. L. 2003, in {\it Modelling of Stellar
Atmospheres} (IAU Symp. 210), eds.  N. Piskunov, W. W. Weiss \& D. F.
Gray, ASP, San Francisco, p. A20

\refb Kroupa P., Gilmore G., Tout Ch. A. 1991, MNRAS, 251, 293

\refb Malkov O. 2002, Ap\&SS, 280, 129

\refb Malkov O. Yu., Mironov A. V., Sichevskij S. G. 2011a, EAS Pub.
Ser. 45, 409

\refb Malkov O. Yu., Mironov A. V., Sichevskij S. G. 2011b, Ap\&SS, 335,
105

\refb Piskunov A. E., Malkov O. Yu. 1991, A\&A, 247, 87

\refb Sichevskij S. G., Mironov A. V., Malkov O. Yu. 2014,
Astrophysical Bulletin, 69, 160

\refb Zakharov A. I., Mironov A. V., Prokhorov M. E. et al. 2013a,
Astronomy Reports, 57, 195

\refb Zakharov A. I., Mironov A. V., Nikolaev F. N. et al. 2013b,
AN, 334, 823

\refb Zakharov A. I., Mironov A. V., Prokhorov M. E. et al. 2013c,
AN, 334, 828

\end{document}